\documentclass[12pt]{article}

\usepackage{amssymb}
\usepackage{amsmath}

\begin{document}%


\title{\bf Approximate formulas for moderately small eikonal amplitudes}

\author{A.V. Kisselev\thanks{Electronic address:
alexandre.kisselev@ihep.ru} \\
{\small Institute for High Energy Physics, NRC ``Kurchatov
Institute''\!,} \\
{\small 142281 Protvino, Russian Federation}}

\date{}

\maketitle

\thispagestyle{empty}

\bigskip

\begin{abstract}
The eikonal approximation for moderately small scattering amplitudes
is considered. With the purpose of using for their numerical
estimations, the formulas are derived which contain no Bessel
functions, and, hence, no rapidly oscillating integrands. To obtain
these formulas, the improper integrals of the first kind which
contain products of the Bessel functions $J_0(z)$ are studied. The
expression with four functions $J_0(z)$ is generalized. The
expressions for the integrals with the product of five and six
Bessel functions $J_0(z)$ are also found. The known formula for the
improper integral with two functions $J_\nu(z)$ is generalized for
non-integer $\nu$.
\end{abstract}




\section{Introduction}

The eikonal approximation was born in the study of a ray optics in
which it is assumed that light travels in a straight line. This
assumption works fine as long as the size of the obstacle $a$ is
large compared to the wavelength of light $\lambda$.

In quantum mechanics the eikonal approximation works well for
processes involving the scattering of particles with large incoming
momentum $k$ and when the scattering angle $\theta$ is very small.
In potential scattering the eikonal approximation may be used if
\cite{Moliere}, \cite{Glauber}
\begin{equation}\label{energy_condidtion}
E  \gg V(\vec{r}) \;,
\end{equation}
where $E$ is the particle energy, and $V(\vec{r})$ is the
interaction potential. The cross section is defined by the amplitude
as
\begin{equation}\label{cross_section_potential}
\frac{d\sigma}{d\Omega} = |A(\vec{k}, \vec{k}')|^2 \;,
\end{equation}
where $\vec{k}$ and $\vec{k}'$ are incoming and outgoing 3-momenta
of a particle. It is  convenient to define a momentum transfer:
\begin{equation}\label{transfer_momentum}
t = -(\vec{k}- \vec{k}')^2 = -2k^2 (1 - \cos \theta) \;.
\end{equation}
The Born scattering amplitude looks like
\begin{equation}\label{Born_amplitude}
A_B(\vec{k}, \vec{k}') = - \frac{1}{4\pi} \!\int \!d\vec{r} \,
\exp[i(\vec{k} - \vec{k}')\vec{r}] V(\vec{r}) \;.
\end{equation}

In eikonal approximation for a spherically symmetric potential the
amplitude can be presented in the form (see, for instance,
\cite{Newton}):
\begin{equation}\label{eikonal_amplitude_potential}
A(\vec{k}, \vec{k}') = i k \int\limits_0^{\infty} \!b \, db \, J_0
(b \sqrt{-t}) \{ 1 - \exp[i\chi(b)] \} \;,
\end{equation}
with the eikonal $\chi(b)$ given by the formula
\begin{equation}\label{eikonal_potential}
\chi(b) = - \frac{k}{E} \int\limits_b^{\infty} \!\frac{r
dr}{\sqrt{r^2 - b^2}} \, V(r) \;.
\end{equation}
Note that the eikonal is the Fourier transform of the Born
amplitude,
\begin{equation}\label{eikonal_Born_amplitude}
\chi(b) = \frac{1}{2\pi k} \int \!d^2k \exp(-i\vec{k}\vec{b})
A_B(\vec{k}, \vec{k}') = \frac{1}{k} \int\limits_0^{\infty} \!dt
J_0(b\sqrt{-t}) A_B(s,t) \;,
\end{equation}
where $s = k^2$.

In perturbative quantum field theory an exponentiation in
high-energy scattering processes similar to that of the Glauber
approximation was studied in \cite{Cheng}. The eikonal
representation can be derived in the framework of quasipotential
approach \cite{Logunov} for small scattering angles and smooth
quasipotentials \cite{Garsevanishvili}. In \cite{Petrov} an
extention of the eikonal approach was developed which automatically
takes into account off-shell unitarity.

The differential cross section is given by
\begin{equation}\label{diff cross sec_def}
\frac{d\sigma(s,t)}{dt} = \frac{1}{16 \pi s^2} \left| A(s, t)
\right|^2 \;,
\end{equation}
where $A(s, t)$ is the (dimensionless) scattering amplitude, with
$s$ and $t=-q_{\bot}^2$ being Mandelstam variables. In the eikonal
approximation the amplitude is defined by the formula
\begin{equation}\label{amplitude}
A(s, t) = 4 \pi i s \!\! \int\limits_0^{\infty} \! db \, b \, J_0(b
\sqrt{-t}) \left\{ 1 - \exp [i \chi(s, b)] \right\} \;,
\end{equation}
where $b$ is the impact parameter. In its turn, the eikonal
$\chi(s,b)$ in \eqref{amplitude} is related to the Born amplitude
$A_B(s,t)$ by the Fourier-Bessel transformation,
\begin{equation}\label{eikonal}
\chi(s, b) = \frac{1}{4\pi s} \int\limits_0^{\infty} \! dq_{\bot}
q_{\bot} J_0(q_{\bot}  b) \, A_{\mathrm{B}} (s, -q_{\bot}^2) \;.
\end{equation}
If $A_{\mathrm{B}}$ is approximated by reggeons, we come to
so-called Regge-eikonal approach.%
\footnote{In the presence of extra dimensions, the Born amplitude
gets an additional contribution from \emph{gravi-reggeons}
(reggeized Kaluza-Klein gravitons in the $t$-channel)
\cite{Kisselev}.}

Suppose, one needs to estimate the amplitude (and, correspondingly,
the differential cross section) numerically at
\begin{equation}\label{eikonal_kinematic}
s \gg -t \gg m_N^2 \;,
\end{equation}
as, for example, it takes place in studying interaction of
high-energy cosmic particles with atmospheric nucleons. As one can
see from eqs.~\eqref{amplitude}, \eqref{eikonal}, the amplitude is
defined via $ A_{\mathrm{B}}$ by iterated integral which contains
\emph{rapidly oscillating} Bessel functions. An attempt to calculate
this integral in the kinematical region \eqref{eikonal_kinematic}
with a high accuracy may face computing difficalties.

The problem becomes much easier if $|\chi(s, b)| \ll 1$ for all $b$
at fixed $s$ under consideration. In such a case, one can use the
expansion
\begin{equation}\label{eikonal_expansion_1}
1 - \exp (i \chi) = - i \chi + \mathrm{O}(\chi^2) \;,
\end{equation}
and get well-known result $A(s,t) \simeq A_{\mathrm{B}}(s,t)$ (see
eq.~\eqref{A_1_final} below). However, there could be that $|\chi(s,
b)|$ \emph{is moderately small} (i.e. $|\chi(s, b)| < 1$). In such a
case, the use of expansion \eqref{eikonal_expansion_1} is not enough
to achieve a required accuracy of numerical calculations, and a few
more terms must be kept in the r.h.s. of
\eqref{eikonal_expansion_1}. This is the goal of the present paper.


\section{Moderately small eikonal amplitude}

In our analysis, we restrict ourselves to three terms in the Taylor
series expansion of $\exp (i \chi)$ in eq.~\eqref{amplitude}:%
\footnote{For more than three terms in the r.h.s. of
\eqref{eikonal_expansion_3}, see our comments in the end of this
Section.}
\begin{equation}\label{eikonal_expansion_3}
1 - \exp (i \chi) = - i \left( \chi - \frac{\chi^3}{6} \right) +
\frac{\chi^2}{2} + \mathrm{O}(\chi^4) \;.
\end{equation}
Then the corresponding approximation for the amplitude looks like
\begin{equation}\label{amplitude_expansion}
A(s, t) \simeq  [A_1(s, t) - A_3(s, t)] + i A_2(s, t)  \;,
\end{equation}
where
\begin{align}
A_1(s, t) &= 4\pi s \int\limits_0^{\infty} \! db \, b \, J_0(b
\sqrt{-t}) \chi(s, b) \;, \label{A_1_def} \\
A_2(s, t) &= 2\pi s \int\limits_0^{\infty} \! db \, b \, J_0(b
\sqrt{-t}) \chi^2(s, b) \;, \label{A_2_def} \\
A_3(s, t) &= \frac{2}{3} \pi s \int\limits_0^{\infty} \! db \, b \,
J_0(b \sqrt{-t}) \chi^3(s, b) \label{A_3_def} \;.
\end{align}
Omitting terms of order $\mathrm{o}(\chi^4)$, we find
\begin{align}\label{diff cross sec}
\frac{d\sigma(s,t)}{dt} = \frac{1}{16\pi s^2} & \{ |A_1(s, t)|^2 - i
[A_1(s, t) A_2^\ast(s, t) - A_1^\ast(s, t) A_2(s, t)]
\nonumber \\
&+ |A_2(s, t)|^2 - [A_1(s, t) A_3^\ast(s, t) + A_1^\ast(s, t) A_3(s,
t] \}\;.
\end{align}
In particular, if $A_B(s,t)$ has no imaginary part, then in the same
approximation the differential cross section looks like
\begin{equation}\label{diff cross sec_real}
\frac{d\sigma(s,t)}{dt} = \frac{1}{16 \pi s^2} \left\{ A_1^2(s, t) +
A_2^2(s, t) - 2A_1(s, t) A_3(s, t) \right\} \;.
\end{equation}
For a pure imaginary Born amplitude, $A_1$ and $A_3$ are pure
imaginary, while $A_2$ is real, and we find:
\begin{align}\label{diff cross sec_real}
\frac{d\sigma(s,t)}{dt} = \frac{1}{16 \pi s^2} &\,\big\{ A_1^2(s, t)
- 2i A_1(s, t) A_2(s, t)
\nonumber \\
&+ A_2^2(s, t) + 2A_1(s, t) A_3(s, t) \big\} \;.
\end{align}

Let us study three parts \eqref{A_1_def}-\eqref{A_3_def} of the
amplitudes separately. Taking into account formulas
\eqref{two_Bessels_def}, \eqref{two_Bessels} presented in
Appendix~A, we obtain immediately from \eqref{A_1_def}:
\begin{align}\label{A_1_final}
A_1(s, t) &= \int\limits_0^{\infty} \! db \, b \, J_0 (b
\sqrt{-t}) \int\limits_0^{\infty} \! dq_{\bot} q_{\bot} J_0(q_{\bot} b) A_B(s,q_{\bot})
\nonumber \\
&= \int\limits_0^{\infty} \! dq_{\bot} q_{\bot} A_B(s, q_{\bot})
\int\limits_0^{\infty} \! db \, b \, J_0(q_{\bot} b) J_0(b
\sqrt{-t}) = A_B(s,\sqrt{-t}) \;.
\end{align}

The second part of the amplitude \eqref{A_2_def} can be analyzed by
the use of eqs.~\eqref{three_Bessels_def}, \eqref{three_Bessels}
from Appendix~A. As a result, we find:
\begin{align}\label{A_2_mod_1}
&A_2(s, t)
\nonumber \\
&= \frac{1}{8\pi s} \int\limits_0^{\infty} \! db \, b \, J_0 (b
\sqrt{-t}) \int\limits_0^{\infty} \! dq_{\bot} q_{\bot} J_0(q_{\bot}
b) A_B(s, q_{\bot}) \int\limits_0^{\infty} \! dq'_{\bot} q'_{\bot}
J_0(q'_{\bot} b)  A_B(s, q'_{\bot})
\nonumber \\
&= \frac{1}{8\pi s} \int\limits_0^{\infty} \! dq_{\bot} q_{\bot}
A_B(s,q_{\bot}) \int\limits_0^{\infty} \! dq'_{\bot} q'_{\bot}
 A_B(s,q'_{\bot}) \int\limits_0^{\infty} \! db \, b \, J_0(q_{\bot} b)
J_0(q'_{\bot} b) J_0(b\sqrt{-t})
\nonumber \\
&= \frac{1}{4\pi^2 s} \int\limits_0^{\infty} \! dq_{\bot} q_{\bot}
A_B(s,q_{\bot}) \int\limits_0^{\infty} \! dq'_{\bot} q'_{\bot}
A_B(s,q'_{\bot}) \,
\theta(\sqrt{-t} - |q_{\bot} - q'_{\bot}|)
\nonumber \\
&\times \theta (q_{\bot} + q'_{\bot} - \sqrt{-t}) \, \frac{1}{
\sqrt{[-t - (q_{\bot} - q'_{\bot})^2] [(q_{\bot} + q'_{\bot})^2 +
t]}} \;.
\end{align}

Let us define new variables
\begin{equation}\label{q_1_q_2}
q_1 = q_{\bot} + q'_{\bot} \;, \quad q_2 = q_{\bot} - q'_{\bot} \;.
\end{equation}
In terms of them $A_2(s, t)$ can be rewritten in the form:
\begin{align}\label{A_2_mod_2}
&A_2(s, t) = \frac{1}{32\pi^2 s} \int\limits_{0}^{\infty} \!\! dq_1
\!\! \int\limits_{-q_1}^{q_1} \!\! dq_2 A_B [s, (q_1 + q_2)/2] \,
A_B[s,(q_1 - q_2)/2]
\nonumber \\
&\times \theta (q_1 - \sqrt{-t}) \, \theta (\sqrt{-t} - |q_2|)\,
\frac{ (q_1^2 - q_2^2)}{\sqrt{(-t - q_2^2)(q_1^2 + t)}}
\nonumber \\
&= \frac{1}{16\pi^2 s} \int\limits_{\sqrt{-t}}^{\infty} \!\! dq_1
\!\! \int\limits_{0}^{\sqrt{-t}} \!\! dq_2  A_B[s, (q_1 + q_2)/2] \,
A_B[s,(q_1 - q_2)/2]
\nonumber \\
&\times \frac{ (q_1^2 - q_2^2)}{\sqrt{(-t - q_2^2)(q_1^2 + t)}} \;.
\end{align}
After change of variables $q_i = x_i \sqrt{-t}$ ($i=1,2$) we come to
the equation
\begin{align}\label{A_2_final}
A_2(s, t) &= \frac{1}{16\pi^2} \left( \frac{-t}{s} \right)
\!\int\limits_{1}^{\infty}dx_1 \!\int\limits_{0}^{1} \!\! dx_2
\frac{ (x_1^2 - x_2^2)}{\sqrt{(x_1^2 - 1)(1 - x_2^2)}}
\nonumber \\
&\times A_B[s, \sqrt{-t}(x_1 + x_2)/2] A_B[s, \sqrt{-t}(x_1 -
x_2)/2] \;.
\end{align}

For the third part of the amplitude \eqref{A_3_def} we get
\begin{align}\label{A_3_mod_1}
A_3(s, t) &= \frac{1}{96\pi^2 s^2} \int\limits_0^{\infty} \! db \, b
\, J_0 (b \sqrt{-t}) \int\limits_0^{\infty} \! dq_{\bot} q_{\bot}
J_0(q_{\bot} b) A_B(s, q_{\bot})
\nonumber \\
&\times \int\limits_0^{\infty} \! dq'_{\bot} q'_{\bot} J_0(q'_{\bot}
b) A_B(s, q'_{\bot}) \int\limits_0^{\infty} \! dq''_{\bot}
q''_{\bot} J_0(q''_{\bot} b) A_B(s, q''_{\bot})
\nonumber \\
& = \frac{1}{96\pi^2 s^2} \int\limits_0^{\infty} \! dq_{\bot}
q_{\bot} A_B (s,q_{\bot}) \!\!\int\limits_0^{\infty} \! dq'_{\bot}
q'_{\bot} A_B(s,q'_{\bot}) \!\!\int\limits_0^{\infty} \! dq''_{\bot}
q''_{\bot} A_B(s,q''_{\bot})
\nonumber \\
&\times \int\limits_0^{\infty} \! db \, b \, J_0(q_{\bot} b)
J_0(q'_{\bot} b) J_0(q''_{\bot} b) J_0(b\sqrt{-t}) \;.
\end{align}
By using eqs.~\eqref{four_Bessels_def}, \eqref{four_Bessels_full}
from Appendix~A, the amplitude $A_3(s, t)$ \eqref{A_3_mod_1} can be
be presented in the form:
\begin{align}\label{A_3_mod_2}
A_3(s, t) &= \frac{1}{96\pi^2} \left( \frac{-t}{s} \right)^2
\!\int\limits_0^{\infty} \! dx A_B(s,\sqrt{-t}x)
\!\!\int\limits_0^{\infty} \! dx' A_B(s, \sqrt{-t}x')
\nonumber \\
&\times \!\!\int\limits_0^{\infty} \! dx'' A_B(s,\sqrt{-t}x'') \,
G(x, x', x'') \;.
\end{align}
Here we introduced the function
\begin{align}\label{G_def}
G(x, x', x'') = \left\{
  \begin{array}{cc}
   \displaystyle{\frac{1}{\pi^2 A} \, K \left( \frac{\sqrt{B}}{A} \right)},
   & A^2 > B \;,
\\ \\
   \displaystyle{\frac{1}{\pi^2 \sqrt{B}} \, K \left( \frac{A}{\sqrt{B}} \right)},
   & 0 \leqslant A^2 < B
\;,
\\ \\
0 \;,  & A^2 < 0 \;,
  \end{array}
\right.
\end{align}
with $K(k)$ being the complete elliptic integral of the first kind
\eqref{elliptic_integral} and notations
\begin{equation}\label{B}
16 A^2 = [(x'' + 1)^2 - (x - x')^2][(x + x')^2 - (x'' - 1)^2] \;,
\quad B = x x' x'' \;.
\end{equation}

Let us define new variables
\begin{equation}\label{x_1_x_2_x_3}
x_1 = x + x' \;, \quad x_2 = x - x' \;, \quad x_3 = x'' \;,
\end{equation}
They run in the region restricted by the following inequalities
\begin{equation}\label{x_1_x_2_x_3_limits}
0\leqslant x_1 \leqslant \infty \;, \quad -x_1 \leqslant x_2
\leqslant x_1 \;, \quad 0\leqslant x_3 \leqslant \infty \;,
\end{equation}
and
\begin{equation}\label{positive_delta}
[(x_3 + 1)^2 - x_2^2][x_1^2 - (x_3 - 1)^2] > 0 \;.
\end{equation}
Then we find from eqs.~\eqref{A_3_mod_2}-\eqref{positive_delta}:
\begin{align}\label{A_3_3}
A_3(s, t) &= \frac{1}{96\pi^2} \left( \frac{-t}{s} \right)^2
\!\int\limits_0^{\infty} \! dx_1 \!\! \int\limits_{-x_1}^{x_1} \!
dx_2 \!\! \int\limits_{0}^{\infty} \! dx_3 \, \theta ( [(x_3 + 1)^2
- x_2^2][x_1^2 - (x_3 - 1)^2] )
\nonumber \\
&\times H(s,t; x_1, x_2, x_3) \;,
\end{align}
where
\begin{align}\label{G_tilde}
H(s,t; x_1, x_2, x_3) &= A_B(s,\sqrt{-t}(x_1 + x_2)/2) \, A_B (s,
\sqrt{-t}(x_1 - x_2)/2)
\nonumber \\
&\times A_B(s,\sqrt{-t}x_3) \, G((x_1 + x_2)/2, (x_1 - x_2)/2, x_3)
\;.
\end{align}

By using formula from Appendix~B, we get
\begin{align}\label{A_3_final}
A_3(s,t) &= \frac{1}{96\pi^2} \left( \frac{-t}{s} \right)^2 \!\Bigg[
\int\limits_0^1 \!dx_1 \!\int\limits_{0}^{x_1} \!dx_2
\!\int\limits_{1-x_1}^{x_1+1} \!\!dx_3 + \int\limits_1^{2} \!dx_1
\!\int\limits_{0}^{1} \!dx_2 \!\int\limits_{0}^{x_1+1} \!dx_3
\nonumber \\
&+ \int\limits_1^{2} \!dx_1 \!\int\limits_{1}^{x_1} \!dx_2
\!\!\int\limits_{|x_2|-1}^{x_1+1} \!\!dx_3 + \int\limits_2^{\infty}
\!dx_1 \int\limits_{0}^{1} \!\!dx_2 \!\!\int\limits_{0}^{x_1+1}
\!dx_3 + \int\limits_2^{\infty} \!dx_1 \!\int\limits_{1}^{x_1}
\!dx_2 \!\!\int\limits_{|x_2|-1}^{x_1+1} \!dx_3 \Bigg]
\nonumber \\
&\times \left[ H(s,t; x_1, x_2, x_3) + H(s,t;x_1, -x_2, x_3) \right]
.
\end{align}

Our formulas \eqref{A_1_final}, \eqref{A_2_final}, \eqref{A_3_final}
and \eqref{diff cross sec} can be applied to numerical calculations
of differential cross sections at given fixed $s$ and $t$, provided
that the Born amplitude $A_B(s,t)$ is known and approximation
\eqref{eikonal_expansion_3} is justified. The advantage of these
integrals lies in the fact that they contain no rapidly oscillating
integrands.%
\footnote{The complete elliptic integral of the first kind $K(k)$ in
\eqref{G_def} is a monotone increasing function of its modulus $k$
($0 \leqslant k <1$).}
Of course, we assume that the Born amplitude has no such
oscillations in variable $t$.

If more terms should be kept in expansion
\eqref{eikonal_expansion_3}, integrals with more than four Bessel
functions $J_0(z)$ have to be used. The corresponding formulas are
presented in Appendix~A (see eqs.~\eqref{five_Bessels_full_1},
\eqref{six_Bessels}).



\section*{Acknowledgements}

The author is indebted to V.A.~Petrov and V.E.~Rochev for valuable
remarks.



\setcounter{equation}{0}
\renewcommand{\theequation}{A.\arabic{equation}}

\section*{Appendix A}
\label{app:A}

In this section we consider improper integrals of the first kind
which contain products of the Bessel functions. The well-known
formula for the integral with two functions $J_\nu(z)$ is
generalized for non-integer $\nu$. The expression for the tabulated
integral with four functions $J_0(z)$ is defined more exactly.
Finally, the analytic expression for the integrals with the product
of five and six Bessel functions $J_0(z)$ are derived.

Let us start from the integral with \emph{two} Bessel functions
$J_n(z)$:
\begin{equation}\label{two_Bessels_def}
F_2^{(n)}(a,b) = \int\limits_0^{\infty} \!\! dx x J_n (ax) J_n (bx)
\;, \quad  n=0,1, \ldots \;.
\end{equation}
For $a, b > 0$, it is known to be (see eq.~(3.108) in
\cite{Jackson}, as well as eq.~6.512.8. in \cite{Gradshteyn})
\begin{equation}\label{two_Bessels}
F_2^{(n)}(a,b) = \frac{1}{a} \, \delta (a - b) \;.
\end{equation}

Let us calculate more general integral containing Bessel functions
with non-iteger index $\nu$ (see the problem 3.16(a) in ref.
\cite{Jackson}),
\begin{equation}\label{two_Bessel_nu_def}
F_2^{(\nu)}(a,b) = \int\limits_0^{\infty} \!\! dx x J_\nu (ax) J_\nu
(bx) \;.
\end{equation}
We start from the Weber's second exponential integral (see, for
instance, eq.~13.31.(1) in \cite{Watson}):
\begin{align}\label{Weber_integral}
I_W(a,b; p) &= \int\limits_0^{\infty} \!dx x \exp(-p^2x^2) J_\nu
(ax) J_\nu (bx)
\nonumber \\
&= \frac{1}{2p^2} \exp[-(a^2+b^2)/4p^2] I_0 \!\left( \frac{ab}{2p^2}
\right) \;,
\end{align}
where $ I_\nu (z)$ is the modified Bessel function of the first
kind. This formula is valid for $|\arg p| < \pi/4$, $\mathrm{Re}\,
\nu > -1$. The parameters $a$ and $b$ can be arbitrary complex
numbers.

Let $p>0$ in what follows. For $|\arg ab| < \pi/2$ one can apply the
asymptotic formula 7.23.(2) from \cite{Watson}:
\begin{equation}\label{small_p}
I_\nu \!\left( \frac{ab}{2p^2} \right)\Bigg|_{p \rightarrow 0} =
\frac{1}{\sqrt{\pi ab/p^2}} \, \exp(ab/2p^2) [1 + \mathrm{O}(p^2)]
\;,
\end{equation}
and obtain
\begin{equation}\label{Weber_integral_small_p}
I_W (a,b; p) \big|_{p \rightarrow 0} =\frac{1}{\sqrt{ab}} \,
\frac{1}{2\sqrt{\pi p^2}} \exp(-(a-b)^2/4p^2) [1 + \mathrm{O}(p^2)]
\;.
\end{equation}
It is known that \cite{Gelfand} ($t>0$)
\begin{equation}\label{delta_fun_1}
\frac{1}{2\sqrt{\pi t}} \, \exp(-x^2/4t) \Big|_{t\rightarrow 0}
\rightarrow \delta (x) \;.
\end{equation}
As a result, we come to a generalization of the formula
\eqref{two_Bessels} for $\mathrm{Re}\, \nu > -1$, $|\arg ab| <
\pi/2$:
\begin{equation}\label{two_Bessels_nu}
F_2^{(\nu)}(a,b) = I_W (a,b; p) |_{p \rightarrow 0} =
\frac{1}{\sqrt{ab}} \, \delta (a - b) \;.
\end{equation}

For $\mathrm{Re}\, \nu > -1$ but positive $a$ and $b$, formula
\eqref{two_Bessels_nu} can be also derived by using one of the
discontinuous Weber-Schafheitlin integrals (see eq.~11.4.41. in
\cite{Abramowitz}, or 6.575.1 in \cite{Gradshteyn} and
eq.~2.12.31.1. in \cite{Prudnikov_vol_2}):
\begin{equation}\label{W-S_integral}
\int\limits_0^{\infty} \!\! dx x^{\mu-\nu+1} J_\mu (ax) J_\nu (bx) =
\left\{
  \begin{array}{cc}
   0 , & 0 < b < a \;, \\ \\
   \displaystyle{\frac{2^{\mu-\nu+1} a^\mu (b^2 - a^2)^{\nu - \mu -1}}{b^\nu
   \Gamma(\nu - \mu)}} \;, & b > a > 0 \;,
  \end{array}
\right.
\end{equation}
where $\mathrm{Re} \nu > \mathrm{Re} \mu > -1$. Let us put $\mu =
\nu -(1 + \lambda)$ in \eqref{W-S_integral}, and take the limit
$\lambda \rightarrow -1$ ($\lambda + 1 > 0$). Then we find for
$a,b>0$, $\mathrm{Re}\, \nu > -1$:
\begin{align}\label{W-S_int_limit_1}
\int\limits_0^{\infty} \!\! dx x J_\nu(ax) J_\nu(bx) &= \lim_{\mu
\rightarrow \nu} \int\limits_0^{\infty} \!\! dx x^{\mu-\nu+1} J_\mu
(ax) J_\nu(bx)
\nonumber \\
&= \left( \frac{a}{b}\right)^{\!\nu} \!\lim_{\lambda \rightarrow -1}
\frac{2^{-\lambda}}{a^{1+\lambda}} \frac{(b^2 -
a^2)_+^{\lambda}}{\Gamma(1+\lambda)}  = \frac{1}{a} \, \delta (a -
b) \;,
\end{align}
where we used the formula \cite{Gelfand}
\begin{equation}\label{delta_fun_2}
\lim_{\lambda = -1} \frac{x_+^\lambda}{\Gamma(\lambda + 1)} =
\delta(x) \;,
\end{equation}

The integral with \emph{three} Bessel functions $J_0(z)$,
\begin{align}\label{three_Bessels_def}
F_3(a,b,c) = \int\limits_0^{\infty} \!\! dx x J_0(ax) J_0(bx)
J_0(cx) \;,
\end{align}
where $a, b, c > 0$, is given by the formula (see, for instance,
eq.~13.46(3) in \cite{Watson}, or eqs.~2.12.42.14. and
eq.~2.12.42.15. in \cite{Prudnikov_vol_2}):
\begin{equation}\label{three_Bessels}
F_3(a,b,c) = \left\{
  \begin{array}{cc}
   \displaystyle{\frac{1}{2\pi \Delta_3}} , & \Delta_3^2 > 0 \;, \\ \\
    0, & \Delta_3^2 < 0 \;,
  \end{array}
\right.
\end{equation}
where
\begin{equation}\label{Delta_3}
16 \Delta_3^2 = [c^2 - (a-b)^2][(a+b)^2 -c^2] \;,
\end{equation}
Note that the integral in \eqref{three_Bessels} is divergent if
$\Delta_3^2 = 0$ (it takes place when one of the parameters is equal
to the sum of the others, say, $c=a+b > 0$).

Is eq.~\eqref{three_Bessels} in agreement with
eq.~\eqref{two_Bessels}? To estimate expression
\eqref{three_Bessels} in the limit $c=0$ (but $a, b > 0$), let us
put $c=\varepsilon \ll 1$. We have
\begin{equation}\label{Delta_3_limit}
\frac{1}{2\pi \Delta_3}\Big|_{c = \varepsilon \ll 1} \simeq
\frac{2}{(a+b)} \frac{1}{\pi \sqrt{\varepsilon^2 - (a-b)^2}} \;.
\end{equation}
Consider the function
\begin{equation}\label{delta-like_fun}
f_\varepsilon (x) = \theta (\varepsilon + x) \, \theta (\varepsilon
- x)  \,\frac{1}{\pi \sqrt{\varepsilon^2 - x^2}} \;.
\end{equation}
We find from \eqref{delta-like_fun} that
\begin{equation}\label{delta-like_fun_prop}
\lim_{\varepsilon \rightarrow 0} \int\limits_\alpha^\beta \! dx
f_\varepsilon (x) = \left\{
  \begin{array}{cc}
    0 \;, & \alpha < \beta < 0 \mathrm{\ or \ } 0 < \alpha < \beta \;, \\ \\
    1 \;, & \alpha < 0 < \beta \;.
  \end{array}
\right.
\end{equation}
Let us show that $f_\varepsilon (x)$ is a delta-like sequence.
Consider the following sequence of antiderivative functions
\begin{equation}\label{}
F_\varepsilon (x) = \int\limits_{-1}^x \! f_\varepsilon (z) \, dz
\;.
\end{equation}
In the limit $\varepsilon \rightarrow 0$ the function $F_\varepsilon
(x)$ tends to a constant value which is equal to zero at $x < 0$ and
to 1 at $x >0$. At the same time, it is \emph{uniformly} bounded in
$\varepsilon$ within each interval. Consequently,
\begin{equation}\label{_fun_limit}
F_\varepsilon (x)|_{\varepsilon \rightarrow 0} = \theta (x) \;,
\end{equation}
and, correspondingly,
\begin{equation}\label{delta-like_fun_limit}
f_\varepsilon (x)|_{\varepsilon \rightarrow 0} = \theta' (x) =
\delta (x) \;.
\end{equation}
As a result, we reproduce eq.~\eqref{two_Bessels} (for $n=0$) from
eq. \eqref{three_Bessels} in the limit $c=0$.

Consider the integral with \emph{four} Bessel functions $J_0(z)$:
\begin{equation}\label{four_Bessels_def}
F_4(a,b,c,d) = \int\limits_0^{\infty} \!\! dx x J_0(ax) J_0(bx)
J_0(cx) J_0(dx) \;,
\end{equation}
with $a, b, c, d > 0$. In \cite{Prudnikov_vol_2} (see
eq.~2.12.44.1.) the following formula is presented:
\begin{equation}\label{four_Bessels}
F_4(a,b,c,d) = \left\{
  \begin{array}{cc}
   \displaystyle{\frac{1}{\pi^2 \Delta_4} \, K \!\left( \frac{\sqrt{abcd}}{\Delta_4} \right)},
   & \Delta_4^2 > abcd \;,
\\ \\
   \displaystyle{\frac{1}{\pi^2 \sqrt{abcd}} \, K \!\left( \frac{\Delta_4}{\sqrt{abcd}} \right)},
   & \Delta_4^2 < abcd
\;,
  \end{array}
\right.
\end{equation}
where
\begin{equation}\label{Delta_4}
16 \Delta_4^2 = (a + b + c - d)(a + b + d - c)(a + c + d - b)(b + c
+ d - a) \;,
\end{equation}
and
\begin{equation}\label{elliptic_integral}
K(k) = \int\limits_0^{\pi/2} \! \frac{d\theta}{\sqrt{1 - k^2 \sin^2
\theta}} =  \int\limits_0^1 \! \frac{dt}{\sqrt{(1 - t^2)(1 - k^2
t^2)}}
\end{equation}
is the complete elliptic integral of the first kind ($0 \leqslant k
< 1$) \cite{Bateman_vol_3}. For $\Delta_4^2 = abcd$ the integral
\eqref{four_Bessels_def} is not defined. Indeed, the function $K(k)$
has the branch point $k=1$, and it diverges at $k \rightarrow 1$ as
\cite{Bateman_vol_3}
\begin{equation}\label{K_asymptotic}
K(k)\big|_{k \rightarrow 1} \simeq \ln (4/\sqrt{1-k^2})\;.
\end{equation}

Unfortunately, eq.~\eqref{four_Bessels} does not cover one important
case $\Delta_4^2 < 0$.%
\footnote{If one applies the expression in the second line of
eq.~\eqref{four_Bessels} to negative $\Delta_4^2$, he gives a wrong
result.}
Let us analyze this case in detail and show that $F_4(a,b,c,d) = 0$
for $\Delta_4^2 < 0$. According to formula 2.12.44.7. from
\cite{Prudnikov_vol_2},
\begin{align}\label{four_Bessels_gen}
A_\alpha^{(n)} (c_1, \ldots , c_n) &= \int\limits_0^{\infty} dx
x^{\alpha - 1} \prod_{k=1}^{n} J_0(c_k x) = \frac{2^{\alpha -
1}}{c_n^\alpha} \frac{\Gamma(\alpha/2)}{\Gamma(1 - \alpha/2)}
\nonumber \\
&\times F_C^{(n-1)} \left( \frac{\alpha}{2}, \frac{\alpha}{2};
\underset{n-1} {\underbrace{1, \ldots 1}}; \frac{c_1^2}{c_n^2},
\ldots \frac{c_{n-1}^2}{c_n^2} \right),
\end{align}
for $0 < \mathrm{Re} \, \alpha < n/2+ 1; c_k > 0, k = 1, \ldots n; \
c_n > c_1 +  \ldots + c_{n-1}$. Here
\begin{align}\label{Lauricella_fun}
&F_C^{(n)} (a,b; c_1, \ldots c_n; z_1, \ldots z_n)
\nonumber \\
&= \sum_{k_1, \ldots k_n =0}^{\infty} \!\! \frac{(a)_{k_1 + \,
\ldots \, k_n} (b)_{k_1 + \, \ldots \, k_n}}{(c_1)_{k_1} \ldots
(c_n)_{k_n}} \frac{z_1^{k_1} \ldots z_n^{k_n}}{k_1! \ldots k_n!}
\end{align}
is the Lauricella function, $(a)_k = \Gamma (a+k)/\Gamma(a)$ being
the Pochhammer symbol. The series in \eqref{Lauricella_fun}
converges if
\begin{equation}\label{convergence}
\sum_{i = 1}^n \sqrt{|z_i|} < 1 \;.
\end{equation}
It follows from \eqref{four_Bessels_gen} that $A_2^{(4)}(c_1, \ldots
, c_4) = 0$ for $c_1, \ldots c_4 >0$, $c_4 > c_1 + c_2 + c_3$.

We need to analyze all possible relations between the parameters
$a,d,c,d$. If all of them all equal, $a=b=c=d$, we get $\Delta_4^2 =
a^4 = abcd$. If they are pairwise equal, say, $a=b$, $c=d$, but $a
\neq c$, we find again that $\Delta_4^2 = a^2 c^2 = abcd$. In both
cases the integral $F_4(a,b,c,d)$ does not exists (see comments
after eq.~\eqref{elliptic_integral}).

If $a=b=c \neq d$, the inequality $\Delta_4^2 < 0$ is realized only
for $d> 3a$ (i.e. for $d > a + b + c$). In the case when only two
parameters are equal, say, $a=b \neq c$, $a \neq d$, $c \neq d$, the
inequality $\Delta_4^2 < 0$ holds if $\max(c,d) > 2a + \min(c,d)$
(for instance, $d > a + b + c$, if $d>c$).

Finally all the parameters can be different. Let, for example, $d$
be the largest one, $d > a, b, c$. If $d < a + b + c$, then
$\Delta_4^2 > 0$, and we come to expressions \eqref{four_Bessels}.
Otherwise, if $ d > a + b + c$, then $\Delta_4^2 < 0$. Putting
$n=4$, $\alpha=2$, $c_1=a, c_2=b, c_3=c$ and $c_4=d$ in
eq.~\eqref{four_Bessels_gen}, and taking into account that
$\Gamma(1-\alpha/2)|_{\alpha\rightarrow 2} \rightarrow
2/(2-\alpha)$, we conclude from all said above that
\begin{equation}\label{Delta_2_negative}
F_4(a,b,c,d) = A_2^{(4)} (a,b,c,d) = 0 \;, \quad \Delta_4^2 < 0 \;.
\end{equation}

The condition \eqref{Delta_2_negative} can be derived in a different
way. Let us use the integral representation for the Bessel functions%
\footnote{This representation was obtained by F.W.~Bessel.}
\cite{Bateman_vol_2}:
\begin{equation}\label{Bessel_integral_rep}
J_n(z) = \frac{1}{2\pi} \int\limits_0^{2\pi} \!\!d\varphi
\exp[i(z\sin\varphi - n\varphi)] \;, \quad n=0,1,2 \ldots
\end{equation} Then we get
\begin{align}\label{F_4_repr_1}
F_4(a,b,c,d) &= \frac{1}{(2\pi)^4} \int\limits_0^{\infty} B dB
\int\limits_0^{2\pi} \!\!d\varphi_1 \int\limits_0^{2\pi}
\!\!d\varphi_2 \int\limits_0^{2\pi} \!\!d\varphi_3
\int\limits_0^{2\pi} \!\!d\varphi_4
\nonumber \\
&\times \exp(iB(a\sin\varphi_1 + b\sin\varphi_2 + c\sin\varphi_3 +
d\sin\varphi_4)
\nonumber \\
&= \frac{1}{(2\pi)^5} \int\limits_0^{2\pi} \!\!d\varphi_1
\int\limits_0^{2\pi} \!\!d\varphi_2 \int\limits_0^{2\pi}
\!\!d\varphi_3 \int\limits_0^{2\pi} \!\!d\varphi_4
\int\limits_0^{\infty} \!d\vec{B}
\nonumber \\
&\times \exp \left[ i\vec{B} (\vec{a}+ \vec{b}+\vec{c}+\vec{d})
\right] \;,
\end{align}
where
\begin{equation}\label{b_vector}
\vec{B} = (0, B) \;,
\end{equation}
and two-dimensional vectors are introduced:
\begin{align}\label{vectors}
\vec{a} &= (a \cos \varphi_1, a \sin \varphi_1) \;, \nonumber \\
\vec{b} &= (b \cos \varphi_2, b \sin \varphi_2) \;, \nonumber \\
\vec{c} &= (c \cos \varphi_3, c \sin \varphi_3) \;, \nonumber \\
\vec{d} &= (d \cos \varphi_4, d \sin \varphi_4) \;.
\end{align}
Since
\begin{align}\label{delta_fun}
&\frac{1}{(2\pi)^2} \int\limits_0^{\infty} \!d\vec{B} \exp \left[
i\vec{B} (\vec{a}+ \vec{b}+\vec{c}+\vec{d}) \right] =  \delta^{(2)}
(\vec{a}+ \vec{b}+\vec{c}+\vec{d})
\nonumber \\
&= \delta (a\sin\varphi_1 + b\sin\varphi_2 + c\sin\varphi_3 +
d\sin\varphi_4)
\nonumber \\
&\times \delta (a\cos\varphi_1 + b\cos\varphi_2 + c\cos\varphi_3 +
d\cos\varphi_4) \;,
\end{align}
we find that
\begin{align}\label{F_4_repr_2}
F_4(a,b,c,d) &= \frac{1}{(2\pi)^3} \int\limits_0^{2\pi}
\!\!d\varphi_1 \int\limits_0^{2\pi} \!\!d\varphi_2
\int\limits_0^{2\pi} \!\!d\varphi_3 \int\limits_0^{2\pi}
\!\!d\varphi_4
\nonumber \\
&\times \delta (a\sin\varphi_1 + b\sin\varphi_2 + c\sin\varphi_3 +
d\sin\varphi_4)
\nonumber \\
&\times \delta (a\cos\varphi_1 + b\cos\varphi_2 + c\cos\varphi_3 +
d\cos\varphi_4) \;.
\end{align}

As it follows from the consideration presented above, $\Delta_4^2 <
0$ is realized if only one of the parameters $a,b,c,d$ is larger
than the sum of all other ones. Without loss of generality, let us
assume that $d>a+b+c$. The integral \eqref{F_4_repr_2} is non-zero
if only
\begin{align}\label{delta_eq}
a\sin\varphi_1 + b\sin\varphi_2 + c\sin\varphi_3 + d\sin\varphi_4 &=
0 \;,
\nonumber \\
a\cos\varphi_1 + b\cos\varphi_2 + c\cos\varphi_3 + d\cos\varphi_4
&=0 \;.
\end{align}
In particular, in means that the following equation
\begin{align}\label{unity_cond}
1 &= \sin^2\!\varphi_4 + \cos^2\!\varphi_4 = \frac{1}{d^2} \, [ a^2
+ b^2 + c^2 + 2ab \cos(\varphi_1 - \varphi_2)
\nonumber \\
&+ 2bc \cos(\varphi_2 - \varphi_3) + 2ca \cos(\varphi_3 -
\varphi_1)] = A(a,b,c) \;.
\end{align}
should be valid. But the evident bound,
\begin{equation}\label{bound}
A(a,b,c) \leqslant \frac{(a + b +c)^2}{d^2} < 1 \;,
\end{equation}
says us that the equations \eqref{delta_eq} cannot be simultaneously
satisfied, and we come again to the condition \eqref{Delta_2_negative}.%
\footnote{In the same way, one can show that $F_3(a,b,c) = 0$ if
$\Delta_3^2 < 0$.}

It is necessary to verify whether the integral $F_4(a,b,c,d)$
\eqref{four_Bessels_def} exists for $\Delta_4^2=0$ (inequalities
$a,b,c,d>0$ are assumed). Note that the equality $\Delta_4^2=0$
holds only if one of the parameters $a,b,c,d$ is equal to the sum of
other three parameters. Let $d=a+b+c$. The asymptotic formula for
the Bessel function \cite{Bateman_vol_2},
\begin{equation}\label{Bessel_asym}
J_0(z)\Big|_{|z| \gg 1} = \sqrt{\frac{2}{\pi z}} \cos \left( z -
\frac{\pi}{4} \right) \left[ 1 + \mathrm{O}(|z|^{-1}) \right] \;,
\quad |\arg z| < \pi \;,
\end{equation}
can be exploited, that results in the following expression ($x>0$)
\begin{align}\label{four_Bessels_asym}
&J_0(ax) J_0(bx) J_0(cx) J_0[(a+b+c)x] \Big|_{x \gg 1}
= \frac{1}{2\pi^2 x^2 \sqrt{abc(a+b+c)}} \nonumber \\
&\times \{ \cos (2ax) + \cos (2bx) + \cos (2cx)
\nonumber \\
& \ \ + \sin [2(a+b)x] + \sin [2(b+c)x] + \sin [2(a+c)x] \} \left[ 1
+ \mathrm{O}(x^{-1}) \right]
\nonumber \\
&=J_{asym} (x)  + \mathrm{O}(x^{-3}) \;.
\end{align}

Then we can write for some fixed $x_0>0$:
\begin{align}\label{four_Bessels_mod}
F_4(a,b,c,d) \Big|_{d=a+b+c} \!&= \int\limits_0^{x_0} \!\! dx x
J_0(ax) J_0(bx) J_0(cx) J_0[(a+b+c)x]
\nonumber \\
&+ \int\limits_{x_0}^{\infty} \!\! dx x \Big\{ J_0(ax) J_0(bx)
J_0(cx) J_0[(a+b+c)x] - J_{asym} (x) \Big\}
\nonumber \\
&+ \int\limits_{x_0}^{\infty} \!\! dx x J_{asym} (x)\;,
\end{align}
The first integral in \eqref{four_Bessels_mod} is obviously
well-defined. The integrand in the second integral decreases rapidly
at $x \rightarrow \infty$ as $\mathrm{O}(x^{-2})$. The third
integral in \eqref{four_Bessels_mod} is a linear combination of sine
integrals
\begin{equation}\label{si}
\mathrm{si} (z) = -\int\limits_z^{\infty} \frac{\sin t}{t} dt \;,
\end{equation}
with $z = 2(a+b)x_0, 2(b+c)x_0, 2(a+c)x_0$, and cosine integrals
\begin{equation}\label{Ci}
\mathrm{Ci} (z) = -\int\limits_z^{\infty} \frac{\cos t}{t} dt \;,
\end{equation}
with $z = 2ax_0, 2bx_0, 2cx_0$. Since both $\mathrm{si} (z)$ and
$\mathrm{Ci} (z)$ are well-defined functions for $z > 0$
\cite{Bateman_vol_2}, we conclude that the integral $F_4(a,b,c,d)$
\eqref{four_Bessels_def} is convergent for $\Delta_4^2=0$.

We find that eq.~\eqref{four_Bessels} should be generalized as
follows ($a,b,c,d > 0$):
\begin{equation}\label{four_Bessels_full}
F_4(a,b,c,d) = \left\{
  \begin{array}{cl}
   \displaystyle{\frac{1}{\pi^2 \Delta_4} \, K \!\left( \frac{\sqrt{abcd}}{\Delta_4} \right)},
   & \Delta_4^2 > abcd \;,
\\ \\
   \displaystyle{\frac{1}{\pi^2 \sqrt{abcd}} \, K \!\left( \frac{\Delta_4}{\sqrt{abcd}} \right)},
   & 0 < \Delta_4^2 < abcd
\;,
\\ \\
   \displaystyle{\frac{1}{2\pi \sqrt{abcd}}} \, & \Delta_4^2 = 0 \;,
\\ \\
   0, & \Delta_4^2 < 0 \;.
  \end{array}
\right.
\end{equation}
We took into account that $K(0) = \pi/2$.

Note that
\begin{equation}\label{Delta_4_m}
16 \Delta_4^2 = [(c+d)^2 - (a-b)^2][(a+b)^2 -(c-d)^2] \;.
\end{equation}
So, $\Delta_4^2 \rightarrow \Delta_3^2$ in the limit $d=0$, and
eq.~\eqref{four_Bessels_full} (see the first line in its r.h.s)
turns into eq.~\eqref{three_Bessels}. Correspondingly, the
expression \eqref{four_Bessels_full} transforms into expression
\eqref{two_Bessels} in the limit $c = d = 0$. To demonstrate this,
it is enough to put $c+d=\varepsilon \ll 1$, and then repeat
arguments used in the derivation of formulas
\eqref{Delta_3_limit}-\eqref{delta-like_fun_limit}.

Thus, we have shown that \eqref{two_Bessels}, \eqref{three_Bessels}
and \eqref{four_Bessels_full} form a self-consistent set of
equations in the sense that
\begin{align}\label{set_equations}
F_4(a,b,c,d)\big|_{d=0} &= F_3(a,b,c) \;,
\nonumber \\
F_4(a,b,c,d)\big|_{c=d=0} &= F_2(a,b) \;,
\nonumber \\
F_3(a,b,c)\big|_{c=0} &= F_2(a,b) \;.
\end{align}
Let us stress that this set of equations implies the relations
between corresponding \emph{analytical expressions} for
$F_4(a,b,c,d)$ \eqref{four_Bessels_full}, $F_3(a,b,c)$
\eqref{three_Bessels}, and $F_2(a,b)$ \eqref{two_Bessels}.%
\footnote{As for integrals \eqref{four_Bessels_def},
\eqref{three_Bessels_def}, \eqref{two_Bessels_def} themselves,
similar relations between them are trivial, taking into account that
$J_0(0) = 1$.}

Consider now the integral with \emph{five} Bessel functions $J_0(z)$
(as before, $a,b,c,d,e > 0$):
\begin{equation}\label{five_Bessels_def}
F_5(a,b,c,d,e) = \int\limits_0^{\infty} \!\! dx x J_0(ax) J_0(bx)
J_0(cx) J_0(dx)  J_0(ex) \;.
\end{equation}
Let us write the identity:
\begin{equation}\label{two_Bessels_rep}
J_0(ax) J_0(bx) = \int\limits_0^{\infty} \!dy \delta (x-y) J_0(ay)
J_0(by) \;.
\end{equation}
Using eq.~\eqref{two_Bessels},
\begin{equation}\label{delta_two_Bessels}
\delta (x-y) = y \!\int\limits_0^{\infty} \!dt \, t J_0(tx) J_0(ty)
\;,
\end{equation}
we obtain from \eqref{two_Bessels_rep}:
\begin{equation}\label{two_Bessels_rep_1}
J_0(ax) J_0(bx) = \int\limits_0^{\infty} \!dt \, t J_0(tx)
\!\int\limits_0^{\infty} \!dy  y J_0(ay) J_0(by) J_0(ty) \;.
\end{equation}
Thus,
\begin{align}\label{five_Bessels_rep_1}
F_5(a,b,c,d,e) &= \int\limits_0^{\infty} \!dt \,t
\!\int\limits_0^{\infty} \!\! dx x J_0(cx) J_0(dx)  J_0(ex) J_0(tx)
\nonumber \\
&\times \int\limits_0^{\infty} \!dy  y J_0(ay) J_0(by) J_0(ty) \;.
\end{align}
As a result, we come to the following expression:
\begin{equation}\label{five_Bessels_full_1}
F_5(a,b,c,d,e) = \int\limits_0^{\infty} \!dt \,t \, F_3(a,b,t) \,
F_4(c,d,e,t) \;,
\end{equation}
where the functions $F_3(a,b,t)$ and $F_4(c,d,e,t)$ are defined by
formulas \eqref{three_Bessels} and \eqref{four_Bessels_full},
respectively.

It is possible to get an expression for $F_5(a,b,c,d,e)$ which is
explicitly symmetric with respect to the parameters $a,b,c,d,e$.
First let us note that $e^2 F_5(a,b,c,d,e) = F_5(\tilde{a},
\tilde{b}, \tilde{c}, \tilde{d},1)$, where%
\footnote{Of course, instead of $e$, one can use any of the
parameters $a,b,c,d$.}
\begin{equation}\label{reduced_parameters}
\tilde{a} = \frac{a}{e} \;, \quad \tilde{b} = \frac{b}{e} \;, \quad
\tilde{c} = \frac{c}{e} \;, \quad \tilde{d} = \frac{d}{e} \;.
\end{equation}
By using equations
\begin{align}\label{two_Bessels_rep_2}
J_0(ax) J_0(bx) &= \int\limits_0^{\infty} \!dt \,t J_0(tx)
\!\int\limits_0^{\infty} \!dy y J_0(ay) J_0(by) J_0(ty) \;,
\nonumber \\
J_0(cx) J_0(dx) &= \int\limits_0^{\infty} \!dq \, q J_0(qx)
\!\int\limits_0^{\infty} \!dz z J_0(cz) J_0(dz) J_0(qz) \;,
\end{align}
we obtain:
\begin{align}\label{five_Bessels_full_2}
F_5(a,b,c,d,e) &= \int\limits_0^{\infty} \!dt \, t \, F_3(a,b,t)
\!\int\limits_0^{\infty} \!dq \, q  F_3(c,d,q) F_3(e,t,q)
\nonumber \\
&= \frac{1}{e^2} \int\limits_0^{\infty} \!dt \, t \,
F_3(\tilde{a},\tilde{b},t) \!\int\limits_0^{\infty} \!dq \, q \,
F_3(\tilde{c},\tilde{d},q) F_3(1,t,q) \;,
\end{align}
which is symmetric in dimensional parameters $\tilde{a}, \tilde{b},
\tilde{c}, \tilde{d}$.

Note that
\begin{equation}\label{two_F_3}
\int\limits_0^{\infty} \!dq \, q  F_3(c,d,q) F_3(e,t,q) =
\int\limits_0^{\infty} \!dq \, q \,\frac{\theta(\Delta_3 (c,d,q))\,
\theta(\Delta_3 (e,t,q))}{\Delta_3 (c,d,q) \, \Delta_3 (e,t,q)} \;.
\end{equation}
With the help of formula 1.2.37.1. from \cite{Prudnikov_vol_1}, one
can show that after integration in variable $q$, the r.h.s of
eq.~\eqref{five_Bessels_full_2} turns into r.h.s of
eq.~\eqref{five_Bessels_full_1}.

Analogously, we find ($a,b,c,d,e,f > 0$):
\begin{align}\label{six_Bessels}
&F_6(a,b,c,d,e,f) = \int\limits_0^{\infty} \!\! dx x J_0(ax) J_0(bx)
J_0(cx) J_0(dx) J_0(ex) J_0(fx)
\nonumber \\
&= \int\limits_0^{\infty} \!dt \,t \,F_3(a,b,t)
\!\int\limits_0^{\infty} \!dq \,q  \,F_3(c,d,q)
\!\int\limits_0^{\infty} \!dp \,p \,F_3(e,f,p) \,F_3(t,q,p)
\nonumber \\
&= \int\limits_0^{\infty} \!dt \,t \,F_4(a,b,c,t) \,F_4(d,e,f,t) \;.
\end{align}
The analytic expressions for $F_3(a_1, a_2, a_3)$ and $F_4(a_1, a_2,
a_3, a_4)$ should be taken from \eqref{three_Bessels} and
\eqref{four_Bessels_full}

By doing in the same way, one can express the integrals
\begin{equation}\label{n_Bessels_def}
F_n(a_1, \ldots, a_n) = \int\limits_0^{\infty} \!\! dx x
\prod_{k=1}^n J_0(a_kx) \;,
\end{equation}
with $n > 6$ and $a_k > 0$, $k=1, \ldots, n$ as a
$(n-3)$-dimensional integral of algebraic functions. Note that
$F_n(a_1, \ldots, a_n) = 0$ if $a_n > a_1 + a_2 + \ldots a_{n-1}$.



\setcounter{equation}{0}
\renewcommand{\theequation}{B.\arabic{equation}}

\section*{Appendix B}
\label{app:B}

In this Appendix we study possible restrictions on variables $x_1$,
$x_2$, $x_3$ in the integral \eqref{A_3_final} which follow from the
inequality \eqref{positive_delta}:
\begin{equation}\label{inequality}
[(x_3 + 1)^2 - x_2^2][x_1^2 - (x_3 - 1)^2] > 0 \;.
\end{equation}
Remember that (see eq.~\eqref{x_1_x_2_x_3_limits})
\begin{equation}\label{x_limits} 0\leqslant x_1 \leqslant
\infty \;, \quad -x_1 \leqslant x_2 \leqslant x_1 \;, \quad
0\leqslant x_3 \leqslant \infty \;.
\end{equation}

If $0 \leqslant x_3 < 1$, we get from \eqref{inequality} the
following inequalities
\begin{equation}\label{x_3<1}
\max(|x_2| - 1, 1 - x_1, 0) \leqslant x_3 < 1 \bigcup \, 0 \leqslant
x_3 \leqslant \min(|x_2| - 1, 1 - x_1) \;.
\end{equation}
Analogously, for $x_3 \geqslant 1$ eq.~\eqref{inequality} means that
\begin{equation}\label{x_3>1}
\max(|x_2| - 1, 1) \leqslant x_3 \leqslant 1 + x_1 \bigcup \, 1 +
x_1 \leqslant x_3 \leqslant |x_2| - 1 \;.
\end{equation}
But since $1+x_1 > 1 - |x_2|$, only the first inequality in
\eqref{x_3>1} remains in this case:
\begin{equation}\label{x_3>1_mod}
\max(|x_2| - 1, 1) \leqslant x_3 \leqslant 1 + x_1 \;.
\end{equation}

Let $H(x_1, x_2, x_3)$ be an integrabel function, then
\begin{align}\label{int_limits}
& \int\limits_0^{\infty} dx_1 \int\limits_{-x_1}^{x_1} dx_2
\!\int\limits_0^{\infty} \!\!dx_3 \, \theta ( [(x_3 + 1)^2 -
x_2^2][x_1^2 - (x_3 - 1)^2] ) \, H(x_1, x_2, x_3)
\nonumber \\
&= \Bigg[ \int\limits_0^1 dx_1 \int\limits_{0}^{x_1} dx_2
\!\int\limits_{1-x_1}^{x_1+1} \!\!dx_3 + \int\limits_1^{2} dx_1
\int\limits_{0}^{1} dx_2 \!\int\limits_{0}^{x_1+1} \!\!dx_3
\nonumber \\
&+ \int\limits_1^{2} dx_1 \int\limits_{1}^{x_1} dx_2
\!\int\limits_{|x_2|-1}^{x_1+1} \!dx_3 + \int\limits_2^{\infty} dx_1
\int\limits_{0}^{1} dx_2 \!\int\limits_{0}^{x_1+1} \!\!dx_3
\nonumber \\
&+ \int\limits_2^{\infty} dx_1 \int\limits_{1}^{x_1} dx_2
\!\int\limits_{|x_2|-1}^{x_1+1} \!\!dx_3 \Bigg] \left[ H(x_1, x_2,
x_3) + H(x_1, -x_2, x_3) \right] .
\end{align}




\end{document}